\documentclass[utf8]{FrontiersinHarvard}
\usepackage{url,hyperref,lineno,microtype,subcaption}
\usepackage[onehalfspacing]{setspace}
\usepackage{xspace}
\usepackage{csquotes}
\usepackage[draft,addedmarkup=uuline]{changes}

\newcommand{\package}[1]{\texttt{#1}\xspace}
\newcommand{\code}[1]{\texttt{#1}\xspace}
\newcommand{\github}{GitHub\xspace}

\newcommand{\sunpy}{SunPy\xspace}
\newcommand{\sunpyproj}{SunPy Project\xspace}
\newcommand{\sunpypkg}{\package{sunpy}}
\newcommand{\sunpycode}[1]{\code{#1}}

\newcommand{\astropypkg}{\package{astropy}}
\newcommand{\aiapypkg}{\package{aiapy}}
\newcommand{\pfsspypkg}{\package{pfsspy}}
\newcommand{\soarpkg}{\package{sunpy\_soar}}

\newcommand{\Fido}{\sunpycode{Fido}}
\newcommand{\hpc}{helioprojective Cartesian\xspace}

\newcommand{\hgs}{Stonyhurst heliographic\xspace}
\newcommand{\hgc}{Carrington heliographic\xspace}
\newcommand{\AR}{active region\xspace}

\def\keyFont{\fontsize{8}{11}\helveticabold}
\def\firstAuthorLast{The SunPy Community {et~al.}}
\def\Authors{
The SunPy Community,
Barnes, Will T.\,$^{1,2,*,\dagger}$,
Christe S.$^{1,\dagger}$,
Freij, N.\,$^{3,4,\dagger}$,
Hayes, Laura A.\,$^{5,\dagger}$,
Stansby, David\,$^{6,\dagger}$,
Ireland, Jack\,$^{1}$,
Mumford, Stuart J.\,$^{8}$,
Ryan, Daniel F.\,$^{7}$,
and Shih, Albert Y.\,$^{1}$
}
\def\Address{
$^{1}$ NASA Goddard Space Flight Center, Greenbelt, MD, USA \\
$^{2}$ American University, Washington, DC, USA\\
$^{3}$ Lockheed Martin Solar and Astrophysics Laboratory, Palo Alto, CA, USA \\
$^{4}$ Bay Area Environmental Research Institute, Moffett Field, CA, USA \\
$^{5}$ ESTEC, European Space Agency, Noordwijk, The Netherlands \\
$^{6}$ Advanced Research Computing Centre, University College London, London, United Kingdom \\
$^{7}$ University of Applied Sciences and Arts Northwest Switzerland, Windisch, Switzerland \\
$^{8}$ Aperio Software Ltd, Leeds, England \\
$^\dagger$ These authors contributed equally to this work and share first authorship
}
\def\corrAuthor{Barnes, Will T.}
\def\corrEmail{will.t.barnes@nasa.gov}

\definechangesauthor[name={Will Barnes}, color=orange]{WTB}
\definechangesauthor[name={Nabil Freij}, color=blue]{NF}

\begin{document}
\onecolumn
\firstpage{1}
\title[]{The SunPy Project: An Interoperable Ecosystem for Solar Data Analysis}
\author[\firstAuthorLast]{\Authors}
\address{}
\correspondance{}
\extraAuth{}
\maketitle

\begin{abstract}
The \sunpyproj is a community of scientists and software developers creating an ecosystem of Python packages for solar physics.
The project includes the \sunpypkg core package as well as a set of affiliated packages.
The \sunpypkg core package provides general purpose tools to access data from different providers, read image and time series data, and transform between commonly used coordinate systems.
Affiliated packages perform more specialized tasks that do not fall within the more general scope of the \sunpypkg core package.
In this article, we give a high-level overview of the \sunpyproj, how it is broader than the \sunpypkg core package, and how the project curates and fosters the affiliated package system.
We demonstrate how components of the \sunpy ecosystem, including \sunpypkg and several affiliated packages, work together to enable multi-instrument data analysis workflows.
We also describe members of the \sunpyproj and how the project interacts with the wider solar physics and scientific Python communities.
Finally, we discuss the future direction and priorities of the \sunpyproj.
\tiny
\keyFont{\section{Keywords:} solar physics, sunpy, data analysis, Python, heliophysics}
\end{abstract}

\section{Introduction}
\label{sec:introduction}

The \sunpyproj is an organization whose mission is to develop and facilitate a high-quality, easy-to-use, community-led, free and open-source solar data analysis ecosystem based on the scientific Python environment.
The vision of the project is to build a diverse and inclusive solar physics and heliophysics community that supports scientific discovery and enables reproducibility through the development of accessible, open-source software \citep{bobra_monica_2020_7020094}.
To achieve this mission and to make this vision a reality, the \sunpyproj maintains and guides the development of a number of Python packages including the \sunpypkg core package, and organizes educational activities around the use of Python for solar-physics research.

As the scientific Python environment matured in the early 2010s \citep{Hunter:2007, harris2020array, 2020SciPy-NMeth}, the development of a Python package devoted to solar physics became viable.
This led to the founding of the \sunpyproj in 2011 by scientists at NASA Goddard Space Flight Center.
The goal of the \sunpyproj at that time was to develop a package that provided the core functionality needed for solar data analysis in Python.
To distinguish the software package from the wider project, this original package is now known as the \sunpypkg core package \citep{sunpy_community2020}.
As the \sunpyproj and \sunpypkg grew, an ecosystem of affiliated packages (see \autoref{ssec:affiliated-packages}) was developed to keep the \sunpypkg core package from becoming too large and difficult to manage.

The \sunpyproj is committed to the principles of open development.
All code is hosted and openly-developed on GitHub\footnote{\url{https://github.com/sunpy/sunpy}} in order to enable anyone to contribute code or provide feedback.
All packages within the \sunpyproj must be under an Open Source Initiative (OSI)\footnote{\url{https://opensource.org}} approved license.
Discussion is hosted on several open communication channels which include weekly community calls, mailing lists, a Discourse forum, and instant messaging via Matrix\footnote{\url{https://matrix.org/}}.
Additionally, the \sunpyproj has a code of conduct\footnote{\url{https://sunpy.org/coc}} to ensure that communication within the project is open, considerate, and respectful.

The aim of this paper is to give a high level description of the \sunpyproj, including its various components, and to describe the direction of the project in the coming years.
\autoref{sec:code} describes the various Python packages that form the project, including both the \sunpypkg core package (\autoref{ssec:the-sunpypkg-core-package}) and the various affiliated packages (\autoref{ssec:affiliated-packages}).
\autoref{sec:people} gives an overview of the roles within the project and describes how to become involved with \sunpyproj.
\autoref{sec:community} describes the various activities of the project within the broader solar physics community.
Finally, \autoref{sec:the-future-of-the-sunpyproj} lays out a vision for the future of the \sunpyproj.

\section{Code}
\label{sec:code}

\subsection{The \sunpypkg core package}
\label{ssec:the-sunpypkg-core-package}

The \sunpypkg package is the central pillar of the \sunpyproj \citep{sunpy_community2020} and provides the fundamental tools for accessing, loading, and interacting with solar physics data in Python.
As we will discuss in \autoref{ssec:affiliated-packages}, \sunpypkg functions as one part of a larger ecosystem of packages for doing solar physics research in Python.
While other packages in the ecosystem may focus on particular analysis techniques or analyzing data from specific instruments, the \sunpypkg ``core'' package is focused on providing general tools for working with solar physics data.
As an example, coordinate transformations between common solar coordinate systems are provided by the \sunpypkg core package because they are needed for the analysis of nearly all solar imaging data and are critical for performing multi-instrument studies.
However, correcting an AIA image to account for instrument degradation would not belong in \sunpypkg because it is specific to data from one instrument.
This allows the \sunpypkg core package to be relatively small in size, thereby assuring its maintainability over time.

The primary components of the \sunpypkg package are described briefly in the following paragraphs.
For a more in-depth description of each of these components, see \citet[Section 4]{sunpy_community2020}.
The full documentation of the \sunpypkg Application Programming Interface (API) is provided in the hosted online documentation\footnote{The \sunpypkg API is fully documented here: \url{https://docs.sunpy.org}}.

\subsubsection{Components of the Core Package}
\label{sssec:components-of-the-core-package}

To search for and download data, \sunpypkg provides the \Fido interface for searching across a variety of data providers (e.g., the Virtual Solar Observatory (VSO)\footnote{\url{https://sdac.virtualsolar.org/cgi/search}}, or the Joint Science Operations Center (JSOC)\footnote{\url{http://jsoc.stanford.edu}}) maintained within the solar community.
Internally, \Fido is both an interface that defines the search API for creating data queries as well as a collection of client classes that provide a translation between this user-facing API and the search parameters accepted by individual data providers.
A complete list of all supported data sources is provided in the documentation for using \Fido\footnote{\url{https://docs.sunpy.org/en/stable/guide/acquiring_data/fido.html}}.
Section 4.1.1 of \citet{sunpy_community2020} also provides a comprehensive discussion of the data sources that \Fido searches by default.
Additionally, \Fido can also be extended to search additional data sources that may not be included in \sunpypkg (e.g., the Solar Orbiter Archive, see \autoref{sssec:current-ecosystem}).
Attributes such as time, wavelength, and instrument name, among others, can be used to filter these search results.
By providing a single interface to many disparate data sources, \sunpypkg, via \Fido, easily enables multi-instrument research workflows.

Once a user has downloaded data, the \code{TimeSeries} and \code{Map} objects can be used to load and visualize time series and two-dimensional image data, respectively.
These objects hold the data alongside the associated metadata in order to perform metadata-aware operations such as concatenation for time series or cropping for image data.
In the case of \code{Map}, a World Coordinate System \citep[WCS, e.g.,][]{greisen_representations_2002} is also constructed from the associated metadata to enable easy mapping between pixel and world coordinates via \astropypkg.
In nearly all cases, solar image data is stored in the FITS format \citep{wells_fits_1981} which has an accompanying well-defined metadata standard \citep{pence_definition_2010}.
The accompanying metadata for each \code{Map} object adheres to this standard.
Solar time series data, however, do not have a standard metadata or file format and are stored in a variety of file formats, including FITS, netCDF, JSON, or plain text.
As such, the metadata associated with each \code{TimeSeries} object is much more sparse compared to \code{Map}, but at minimum will include the time of each observation as well as some information about the associated instrument that made the observation.

Additionally, by extending the \astropypkg coordinates framework \citep[see Section 3.3 of][for more details]{the_astropy_collaboration_astropy_2018}, \sunpypkg provides definitions of, and transformations between, common solar coordinate systems.
Coordinates expressed using these frames can be used to represent the positional information of solar features and events. \sunpypkg implements both observer-dependent (e.g., \hpc) and observer-independent (e.g., \hgs) coordinate frames \citep{thompson_coordinate_2006}.
Each \code{Map} object instance also carries with it the corresponding coordinate frame of that image and the coordinate of the observer as defined by the position of the observatory given in the associated metadata.

\subsubsection{Testing Infrastructure}
\label{sssec:testing-infrastructure}

\sunpypkg includes thousands of unit, regression and integration tests that are run using the \code{pytest} testing framework.
This test suite is run on every pull request opened on the \sunpypkg \github repository using \github Actions to ensure that contributions to the codebase do not lead to unexpected regressions.
A full description of our testing practices can be found in our developer documentation.\footnote{A complete guide to running the tests and the associated infrastructure can be found here:\url{https://docs.sunpy.org/en/latest/dev_guide/contents/tests.html}}

\subsubsection{Release Schedule}
\label{sssec:release-schedule}

There is a new release of the core package with feature enhancements approximately every six months.
Every other release is designated a long term support (LTS) release and receives bug fixes for a year rather than for six months.
Additionally, there are bug fix releases every month.
For each release a digital object identifier (DOI) is automatically generated and a record is created on Zenodo.\footnote{The most current release on Zenodo can be found here: \url{https://doi.org/10.5281/zenodo.7314636}}
By providing regularly scheduled, versioned releases of \sunpypkg, the \sunpyproj enables reproducibility.
For example, if a researcher is attempting to reproduce a result from a paper that used \sunpypkg v2.0.2, she can create a new virtual environment and install that exact version of \sunpypkg, even if the current version is many versions ahead of v2.0.2.

This release process is completely automated through \github Actions.\footnote{The \github Actions templates used are available here: \url{https://github.com/OpenAstronomy/github-actions-workflows}}
When a release is tagged, an action is triggered that tests the package on all supported versions of Python and all supported operating systems.
If the packages build successfully, they are automatically uploaded to to Python Package Index (PyPi) and subsequently the release is updated on \code{conda-forge}.

\subsection{Affiliated Packages}
\label{ssec:affiliated-packages}

As the \sunpypkg package grew and the amount of domain- and instrument-specific code being developed in Python increased, it became increasingly challenging to store and maintain the functionality needed for all solar physics research in one package.
As such, the affiliated package system was introduced \citep{mumford_stuart_2014_3261752} so that the \sunpypkg core package could be generic enough for other packages to build on.
The goal of this system is to support and promote software packages outside the scope of the \sunpypkg core package, and to provide guidance to developers in implementing and maintaining the specific functionality provided by an affiliated package.
This fosters code-ownership while ensuring the set of affiliated packages are interoperable and follow a set of common standards (see \autoref{sssec:application-process}).
The \sunpyproj provides development support through our community development efforts and by providing a package template as a foundation.
In addition, affiliated packages are advertised at conferences and workshops where a \sunpy poster, talk, or tutorial is given.

As a result of the creation of the affiliated package ecosystem, components of the \sunpypkg core package that were tied directly to specific instruments or data analysis methods have recently been moved out into other affiliated packages.
One example of this is \aiapypkg \citep{barnes_aiapy_2020}, a package for processing data from the Atmospheric Imaging Assembly \citep[AIA,][]{lemen_atmospheric_2012} on the \textit{Solar Dynamics Observatory} \citep[SDO,][]{pesnell_solar_2012}.
Prior to version 2.1, \sunpypkg included functionality for calibrating level 1 AIA data.
In 2019, in collaboration with the \sunpyproj, the AIA instrument team began developing \aiapypkg to provide a number of AIA-specific analysis routines in Python, including the aforementioned calibration software.
\aiapypkg became an affiliated package in 2020 and the AIA-specific functionality that previously lived in \sunpypkg was deprecated and subsequently removed.
This relocation of the code allows the AIA instrument team to have full autonomy over their calibration routines and release updates to their software on a more frequent timescale than that of the \sunpypkg core package.
At the same time, \aiapypkg users and developers are able to take full advantage of the \sunpyproj ecosystem.

Outside of the current list of affiliated packages, current and future NASA and ESA missions \footnote{This includes, but is not limited to, the Interface Region Imaging Spectrometer (IRIS), several instruments on \textit{Solar Orbiter}, as well as the X-Ray Telescope (XRT) and the Extreme ultraviolet Imaging Spectrometer (EIS) onboard \textit{Hinode}}, as well as ground-based telescopes, such as the Daniel K. Inouye Solar Telescope (DKIST), have begun developing user tools for data analysis and/or pipelines for data calibration built on top of the \sunpy ecosystem.
While these packages are not yet affiliated, the \sunpyproj has assisted in coordinating development efforts between these teams in order to foster a more interoperable ecosystem.

\subsubsection{Application Process}
\label{sssec:application-process}

The affiliated package application process is completed in the open on GitHub and is open to all, both individuals and larger collaborations (e.g., instrument teams).
To begin the process, an applicant opens an issue on the \sunpyproj website GitHub repository\footnote{\url{https://github.com/sunpy/sunpy.org}} and provides details about the package, including the package name, the maintainers, a link to the code repository, and a link to the documentation.
The Affiliated Package Liaison (see \autoref{ssec:community-roles}) then selects a \sunpyproj member to review the candidate affiliated package against the following criteria:

\begin{itemize}
    \item \textit{functionality} --- is the package relevant to the solar physics community?
    \item \textit{integration} --- does the package make use of the existing ecosystem?
    \item \textit{documentation} --- is there hosted documentation, including examples and an API reference?
    \item \textit{testing} --- are there automatically run tests and is the coverage extensive?
    \item \textit{duplication} --- does the package duplicate existing functionality in the ecosystem?
    \item \textit{community} --- is there a code of conduct and do the developers engage the wider community?
    \item \textit{development status} --- is the project actively maintained, including versioned releases?
\end{itemize}

The assigned project member then scores the package in each category using a \enquote{stoplight} system (i.e., a package is scored green, orange, or red in each category).
A detailed description of each criterion and the scoring for each is available on the affiliated package page of the \sunpyproj website\footnote{\url{https://sunpy.org/project/affiliated}}.
The submitting author of the affiliated package may also request an alternate reviewer, in which case the Affiliated Package Liaison will assign a new \sunpyproj member to review the package.
At the end of the review, the candidate package is either accepted, marked as provisional, or not accepted.
If the package is accepted, the affiliated package is added to the list of affiliated packages on the \sunpyproj website.
If the package is marked as provisional or is not accepted, the reviewer and the Affiliated Package Liaison will work with the package authors to help them achieve provisional or accepted status.
Accepted affiliated packages are reviewed once a year to ensure the interoperability of the ecosystem does not regress and that affiliated packages are actively maintained.

In all cases, the goal of the affiliated package review process is to broaden the ecosystem of tools for solar data analysis in Python.
These criteria are not meant to be exclusionary, but rather to ensure interoperability and consistency across the ecosystem for the benefit of both users and developers.
Interoperability in this context, means that affiliated packages should make use of the existing \sunpypkg core data structures, (e.g., \code{Map} and \code{Timeseries}), in lieu of their own custom data structures.
In the context of searching for and downloading data, affiliated packages should use the \Fido interface and extend \Fido for additional data sources as needed.

\subsubsection{Current Ecosystem}
\label{sssec:current-ecosystem}

At the time of writing, the \sunpyproj has a rich and growing ecosystem of affiliated packages.
In addition to the \sunpypkg core package, the affiliated package ecosystem includes:

\begin{itemize}
    \item \aiapypkg for functionality specific to the AIA instrument \citep{barnes_aiapy_2020}
    \item \package{ndcube} for generic handling of $N$-dimensional data sets with a world coordinate system (WCS) \citep{danryanirish_2021_5715161}.
    \item \package{pfsspy} for magnetic-field extrapolation \citep{stansby_pfsspy_2020}
    \item \package{sunkit-instruments} for instrument-specific code that does not have a dedicated package \citep{danryanirish_2022_7190661}.
    \item \package{sunkit-image} for solar-specific image analysis or reduction techniques \citep{nabil_freij_2022_6578722}.
    \item \package{sunpy-soar}\footnote{\url{https://github.com/sunpy/sunpy-soar}} for querying the Solar Orbiter Archive (SOAR)\footnote{\url{https://soar.esac.esa.int/soar/}}
\end{itemize}

\begin{figure}
    \centering
    \includegraphics[width=\columnwidth]{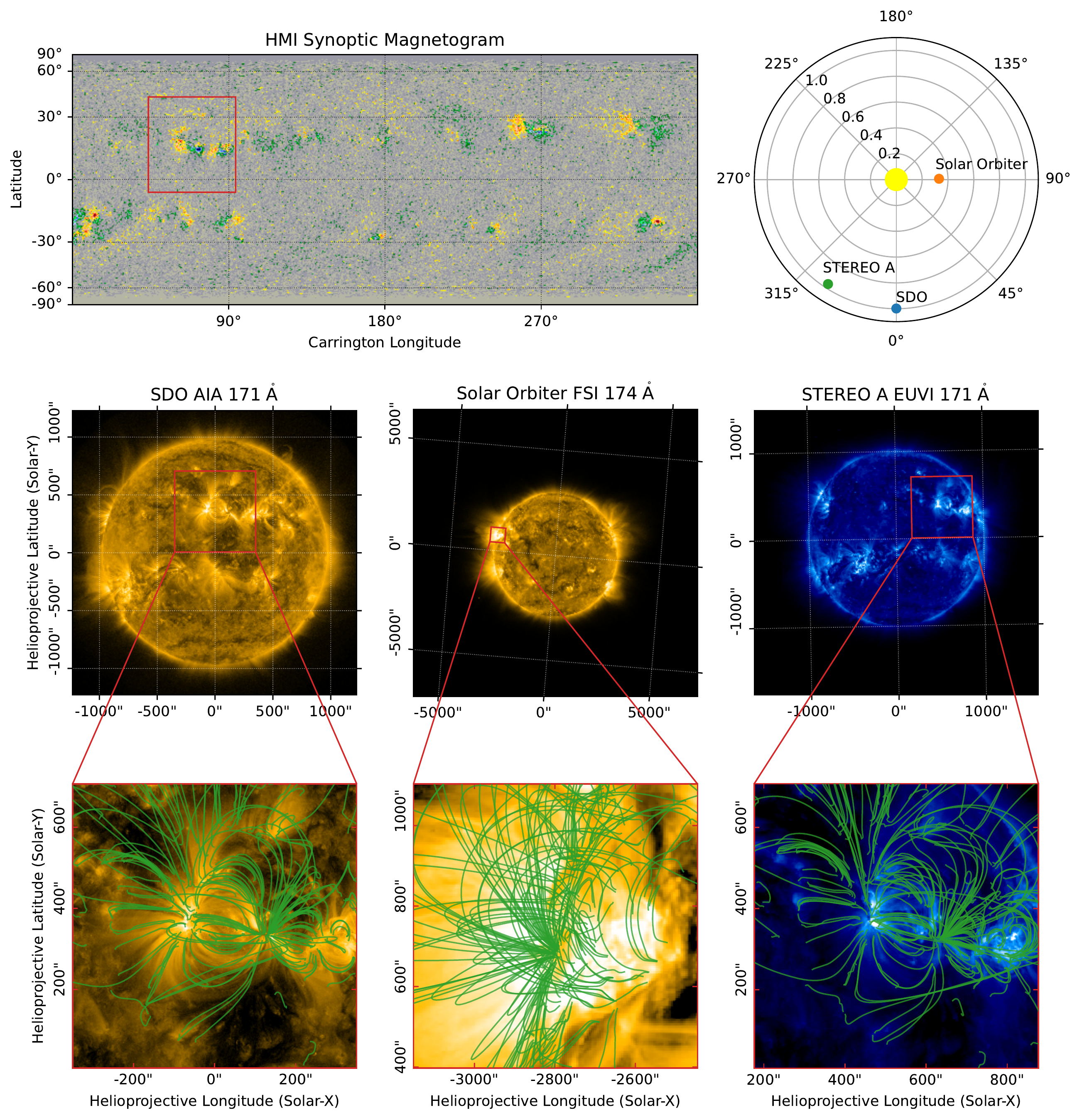}
    \caption{
        Illustration of multiple affiliated packages, including \sunpypkg, \soarpkg, \aiapypkg, and \pfsspypkg, working together.
        \textbf{Top row:} The left panel shows the HMI synoptic magnetogram for Carrington rotation 2255. The red box is centered on the \AR. The right panel shows the \hgs longitude and radius (in AU) for SDO, STEREO A, and \textit{Solar Orbiter} on 2022-03-29.
        \textbf{Middle row:} Full disk images from SDO AIA at 171 Å (left), SolO FSI at 174 Å (middle), and STEREO-A EUVI at 171 Å (right). All three images were downloaded using \sunpypkg along with \soarpkg to query the SOAR for the \textit{Solar Orbiter} image. The AIA image was calibrated using \aiapypkg.
        The red box in each panel is centered on the AR shown in the top panel.
        \textbf{Bottom row:} Cutouts of the regions denoted in each image in the middle row.
        \pfsspypkg is used to compute a potential magnetic field solution from the magnetogram (top row) and trace field lines through the resulting volume.
        These field lines, shown in green, are transformed to the appropriate coordinate system of each instrument using \sunpypkg.
    }
    \label{fig:affiliated-package-showcase}
\end{figure}

To demonstrate how several of the affiliated packages can be used together with \sunpypkg in a scientific workflow, we show an example in \autoref{fig:affiliated-package-showcase} of how coronal loop structures can be analyzed using potential magnetic field extrapolations and multi-point extreme ultraviolet (EUV) observations.
We have included a Jupyter notebook that illustrates each step of this workflow in the \github repository that accompanies this paper.\footnote{The \github repository for this paper, including the complete text and all code to generate \autoref{fig:affiliated-package-showcase}, can be found at \url{https://github.com/sunpy/sunpy-frontiers-paper}}

First, we use the \Fido interface provided by \sunpypkg to search for and download a synoptic magnetogram from the Helioseismic Magnetic Imager \citep[HMI,][]{scherrer_helioseismic_2012} on SDO for Carrington rotation 2255 which began on 2022-03-08.
This is shown in the left panel in the top row of \autoref{fig:affiliated-package-showcase}.
Next, we identify \AR NOAA 12976 which appeared near disk center, as seen from SDO, at 2022-03-29 21:04.
The red box overlaid on the synoptic magnetogram is centered on the \AR when it appeared at disk center at a Carrington longitude of $65^\circ$.

Since we are interested in the EUV observations of \AR 12976, we also use \Fido to query the VSO for data from AIA on SDO and the Extreme Ultraviolet Imager (EUVI) on the \textit{Solar Terrestrial Relations Observatory} \citep[STEREO,][]{howard_sun_2008}.
Additionally, we use the \package{sunpy-soar} package to allow \Fido to search for and download data from the SOAR.
Here, we query the SOAR for data from the Extreme Ultraviolet Imager \citep[EUI,][]{rochus_solar_2020} on \textit{Solar Orbiter} \citep{muller_solar_2020}.

The middle row of \autoref{fig:affiliated-package-showcase} shows full disk EUV images from AIA (left), the full-sun imager (FSI) on EUI (middle), and EUVI on the STEREO-A spacecraft (right).
We use the \aiapypkg package to correct the AIA image (middle panel) for instrument degradation and update the pointing information.
The red box in each panel is centered on \AR 12976, as seen from the respective spacecraft, and has a width and height of 700 arcseconds.
The top right panel of \autoref{fig:affiliated-package-showcase} shows the \hgs longitude and radius (in AU) of SDO, STEREO-A and \textit{Solar Orbiter} as derived from the observer location metadata of each image.

Viewing the \AR from the vantage points of these three spacecraft (separated by $>90^{\circ}$), we gain a better understanding of its three-dimensional structure.
Additionally, we use the \package{pfsspy} package to compute a potential field extrapolation from the corresponding synoptic magnetogram as shown in the top row of \autoref{fig:affiliated-package-showcase}.
We trace field lines from areas of negative magnetic flux inside the red box corresponding to \AR 12976.
The resulting field lines are overlaid in green on top of the cutouts from each EUV image in the bottom row of \autoref{fig:affiliated-package-showcase}.
Each field line traced using \package{pfsspy} is an \astropypkg coordinate object expressed in terms of a \hgc coordinate frame defined in \sunpypkg.
As such, it is straightforward to transform each field line to the observer-dependent coordinate frame of each image as defined by corresponding observatory using the plotting functionality provided in \astropypkg.
The interoperability between \astropypkg, \sunpypkg, \package{sunpy-soar}, \aiapypkg, and \package{pfsspy} allows us to easily examine the three-dimensional magnetic structure of the \AR and see to what extent the derived potential field corresponds to the EUV emission as observed by these three spacecraft.

\section{People}
\label{sec:people}

\subsection{Board and Lead Developers}
\label{ssec:board-and-lead-developers}

The current structure of the \sunpyproj is governed by the \sunpyproj board \citep{christe_steven_2018_3261663}.
The board is a self-electing oversight board which delegates the majority of the day-to-day operations of the project to a lead developer, who in turn delegates it to members of the community.
The lead developer has overall responsibility for the large scale organization of the \sunpypkg core package, and ensures that pull requests comply with stated standards and align with the goals of the \sunpyproj.
The deputy lead developer supports the lead developer and fills in when the lead is absent.
The board's role is to steer the overall direction of the \sunpyproj and consists of scientists and researchers who are not necessarily involved directly with the day-to-day development of the \sunpypkg core package.

\subsection{Community Roles}
\label{ssec:community-roles}

There are several specific community (or executive) roles within the \sunpyproj that perform important duties related to the overall development and maintenance of the project. These roles encompass a range of responsibilities from the development of the core package and affiliated packages to project communication and liaison. The community roles are held by members of the wider solar community who are actively involved in the \sunpyproj. Anyone interested in a community role is encouraged to apply.

At present, there are nine community roles within the \sunpyproj.
From the development side, these include the Lead Developer and the Deputy Lead Developer who are responsible for the development of the \sunpypkg core package, support the development of affiliated packages, and lead the development community.
To assist the Lead/Deputy Developers, there are several development community roles which include:

\begin{itemize}
    \item Continuous Integration Maintainer
    \item Release Manager
    \item Webmaster
    \item Communication and Education Lead who is responsible for the overall engagement with the wider community
    \item Lead Newcomer and Summer of Code mentor who assists new contributors and oversees the Google Summer of Code project
    \item Affiliated Package Liaison who is responsible for overseeing the affiliated package review process (see \autoref{sssec:application-process}) and working with developers of current and candidate affiliated packages
\end{itemize}

\subsection{Maintainers and Contributors}
\label{ssec:maintainers-and-contributors}

The development of the \sunpypkg core package depends principally on an established team of volunteers that support the Lead and Deputy Lead Developers.
These volunteer \textit{maintainers} are given commit access to the \sunpypkg repository and are predominantly, though not exclusively, scientists from the solar community who use \sunpypkg in their work.
In addition to this group of core maintainers, there is a steady influx of new \textit{contributors}, averaging around 20--25 people per year.
These contributors enable a wider range of features and code changes than would otherwise be normally possible due to the time constraints of the established team of volunteers.
Within this subset of maintainers are members who maintain the specific sub-packages within \sunpypkg like \code{sunpy.map} or \code{sunpy.coordinates}.
These individuals are selected due to either their specific knowledge of the topic or their expertise with these sub-packages.

Contributing to the \sunpyproj includes a wide range of activities, not all of them programming related.
These include reporting bugs by raising issues on GitHub, requesting features, writing code and tests, providing feedback on pull requests, correcting or adding documentation, helping people who have problems or questions in communication channels and more.
The \sunpyproj is always looking for any new volunteers or people willing to contribute their time.

\section{Community}
\label{sec:community}

\subsection{Engagement with the Solar Physics Community}
\label{ssec:engagement-with-the-solar-physics-community}

In order for the \sunpyproj to maintain and grow the \sunpypkg core package and affiliated packages within the ecosystem, engagement with the wider solar physics community is critical.
The mission of the \sunpyproj is to be community-led, and the development is driven by the needs of the solar physics community.
To facilitate this, the \sunpyproj is building a community for which there is inclusive and open communication between those developing \sunpypkg and those using \sunpypkg in their scientific research.
Active contributions from users in terms of bug reports, issues encountered with code or documentation, and feature requests are all vital to the sustainability and future of the \sunpyproj.
We emphasize that being part of the \sunpyproj does not necessarily mean writing software. Contributions in the form of feedback and suggestions are equally important.

To foster communication, the \sunpyproj supports several communication platforms (see \autoref{sssec:communication-channels}) through which users and developers can regularly interact.
The \sunpyproj posts on solar physics noticeboards about recent releases and regularly advertises \sunpypkg and affiliated packages at scientific conferences, providing tutorials and support.
We also ask that if \sunpypkg is used for scientific work that it is cited in the literature\footnote{See this page for a guide on how to cite \sunpypkg in published works: \url{https://sunpy.org/about\#acknowledging-or-citing-sunpy}}, thereby increasing its visibility to the scientific community and ultimately contributing to the continued growth and development of the package.
More recently, the \sunpyproj has improved communications and established relationships with data providers such as VSO and the SOAR, and teams supporting both operating and developing instruments and missions.
The \sunpyproj is always looking for ways to improve the accessibility of the project and to grow the community.

\subsubsection{Communication Channels}
\label{sssec:communication-channels}

Over the years, the usage of \sunpypkg and affiliated packages within the solar physics community has increased, and with that methods to communicate within the SunPy community have also increased.
At the time of writing, several distinct communication channels are available.
These include:

\begin{itemize}
    \item Multiple GitHub repositories for bug reports and feature requests.
          These are listed under the \sunpyproj GitHub organization \footnote{\url{https://github.com/sunpy}}.
    \item Real time messaging \footnote{Links to join the Matrix chat can be found at \url{https://sunpy.org/help.html}}.
    \item Mailing lists \footnote{\url{https://groups.google.com/g/sunpy}}.
    \item An online community forum \footnote{\url{https://community.openastronomy.org/}}.
    \item Weekly public calls that anyone can participate in \footnote{\url{https://sunpy.org/jitsi}}.
\end{itemize}

Each has their own distinct purpose, and was created as a need arose for their existence.
For example, the GitHub repository is used for the development of \sunpypkg and issues and bugs can be raised there.
However some scientists may not be familiar with GitHub and would like to ask a general question on how to do something within \sunpypkg.
For this, the mailing list, community forum, or real time Matrix chat may be the most appropriate.
We actively encourage users and those interested in contributing to use any or all of these communication channels.

In addition to the main communication platforms specific to the \sunpyproj, we maintain a presence within other communication channels used by the wider heliophysics community, including Helionauts\footnote{\url{https://helionauts.org/}} and communication channels used by the Python in Heliophysics Community.

\subsection{Python in Heliophysics Community (PyHC)}
\label{ssec:python-in-heliophysics-community-pyhc}

The Python in Heliophysics Community (PyHC)\footnote{\url{https://heliopython.org/}} \citep{barnum2022python} is a project with similar goals as the \sunpyproj, but focuses on the wider Heliophysics community \citep{https://doi.org/10.1029/2018JA025877}.
These include providing coding standards \citep{annex_a_2018_2529131}, curating a list of participating projects\footnote{\url{https://heliopython.org/projects/}}, hosting bi-monthly community meetings, and organizing an inaugural summer school for early career researchers.
The \sunpyproj is actively involved in PyHC, with \sunpypkg being one of the core PyHC packages.
\sunpyproj members regularly attend community meetings and present updates.
The \sunpyproj also took part in the PyHC 2022 summer school.
Moving forward, PyHC and the \sunpyproj will continue to collaborate and build upon efforts of using \sunpypkg and affiliated packages within the larger heliophysics Python ecosystem.

\subsection{Collaboration with the Wider Python Ecosystem}
\label{ssec:collaboration-with-the-wider-python-ecosystem}

The \sunpypkg package forms part of the wider Python scientific ecosystem, requiring active collaboration with other scientific Python packages.
Whenever possible, we aim to contribute to relevant open-source projects rather than duplicating functionality.
As an example, large parts of \sunpypkg depend on core functionality developed in the \astropypkg package, including support for handling units, times, and coordinates.

The \sunpyproj is sponsored by NumFOCUS, \enquote{a nonprofit supporting open code for better science}\footnote{\url{https://numfocus.org/}}.
NumFOCUS provides financial and organizational support for several important packages (e.g., \package{numpy}, \package{pandas} and \package{xarray}) and facilitates collaboration between packages throughout the scientific Python ecosystem.
One example of this is the annual NumFOCUS summit that brings together the leaders of these packages to discuss interoperability, funding sources and other high-level topics that improve the Python ecosystem as a whole.

In addition, the \sunpyproj is a member of the OpenAstronomy organization\footnote{\url{https://openastronomy.org/}}.
OpenAstronomy was created to collaborate on outreach, organize conferences such as Python in Astronomy, develop common tooling for infrastructure, and apply to internship programs such as Google Summer of Code (GSoC)\footnote{\url{https://summerofcode.withgoogle.com/}} and Outreachy\footnote{\url{https://www.outreachy.org/}}.
GSoC has been an invaluable source of programming effort for the \sunpyproj over the past decade.
The contributions from participants in this program have been crucial to \sunpypkg.
Examples of successful projects include the conversion to using \package{astropy.time} and creating a new Python API wrapper for Helioviewer.org\footnote{\url{https://hvpy.readthedocs.io/}}.
As the focus of the OpenAstronomy organization is the broader astrophysics and astronomy community, the \sunpyproj's participation has enabled closer ties within the rapidly growing  Python-in-astronomy landscape.

\section{The Future of The \sunpyproj}
\label{sec:the-future-of-the-sunpyproj}

Development within the \sunpyproj is driven by, and for, the solar physics community, responding to the needs of researchers for data analysis tools and techniques, and software for working with data from new missions.
This means both the \sunpypkg core package and other affiliated packages are continually changing and expanding.
In September 2022 several members of the \sunpyproj met at a coordination meeting to discuss the future of the project.
Two key areas that emerged were updating the governance structure, and creating a roadmap for future development.
The roadmap provides:
\begin{enumerate}
    \item a set of priorities for developers to work on in the medium term.
    \item a well scoped list of work items that funding can be sought for.
    \item a mechanism to solicit input from the wider solar physics community on the medium term priorities from \sunpyproj.
\end{enumerate}

At the time of writing, items in the draft roadmap include:
\begin{itemize}
    \item Improving support and functionality for data with spectral coordinates (e.g., rastering spectrometers)
    \item Enabling multi-dimensional data sets (i.e., beyond 2D images).
    \item Improving support for running \sunpypkg on cloud infrastructure.
    \item Creating a consistently structured set of documentation across all the \sunpyproj packages.
    \item Adding functionality to rapidly visualize large data sets.
    \item Restructuring the project governance to, among other things, transform the lead developer positions into a multi-person steering committee and create an ombudsperson role.
\end{itemize}

The next step is consultation with the wider solar physics community.
We invite feedback on this proposed roadmap via any of the aforementioned communication channels (see \autoref{sssec:communication-channels}) or by opening an issue on the repository used for tracking high-level, project-wide tasks and suggestions\footnote{\url{https://github.com/sunpy/sunpy-project}}.

\section{Conclusion}
\label{sec:conclusion}

In this paper, we have summarized the \sunpyproj and its various components, including the code developed and maintained by the project (\autoref{sec:code}), the people that comprise the project (\autoref{sec:people}) and the community that the project serves (\autoref{sec:community}).
In particular, we have discussed how the \sunpypkg package and the wider set of affiliated packages form a software ecosystem for solar physics research in Python and illustrated the types of analyses that such an ecosystem enables (see \autoref{fig:affiliated-package-showcase}).
Finally, we have summarized a tentative roadmap to steer the direction of the \sunpyproj in the coming years.
Importantly, we hope that such a high level description will provide a more clear understanding of the \sunpyproj and the wider ecosystem and will encourage contributions of all forms from the global solar physics community.

\section*{Appendix}
\label{sec:appendix}

Here we provide a glossary of terms used throughout this paper:

\begin{itemize}
\item \sunpyproj: The board and lead/deputy developers, the community roles, maintainers, and every package under its supervision.
\item \sunpypkg: The core package for using Python for scientific research in solar physics.
\item SunPy ecosystem: The collection of packages that use or interface with \sunpypkg and support scientific research in solar physics, including \sunpypkg
\item Affiliated package(s): Solar physics related functionality outside the scope of the \sunpypkg core package and that satisfies the standards enumerated in \autoref{sssec:application-process}.
\end{itemize}

\section*{Conflict of Interest Statement}
\label{sec:conflict-of-interest-statement}

The authors declare that the research was conducted in the absence of any commercial or financial relationships that could be construed as a potential conflict of interest.

\section*{Author Contributions}
\label{sec:author-contributions}

W.T.B. contributed to writing the text and created \autoref{fig:affiliated-package-showcase}.
S.C., N.F., L.A.H, and D.S. contributed to writing the text.
All authors contributed to manuscript planning and revision and have read and approved the submitted version.

\section*{Data Availability Statement}
\label{sec:data-availability-statement}

All of the data used to create \autoref{fig:affiliated-package-showcase} are publicly available at the following repositories:

\begin{itemize}
    \item AIA and EUVI data are available through the VSO: \url{https://sdac.virtualsolar.org/cgi/search}
    \item EUI data are available through the SOAR: \url{https://soar.esac.esa.int/soar/}
    \item HMI synoptic magnetogram data are available through the JSOC: \url{http://jsoc.stanford.edu/}
\end{itemize}

Furthermore, the text and accompanying scripts to query, download, and process the data and make \autoref{fig:affiliated-package-showcase} are publicly available in the \github repository\footnote{\url{https://github.com/sunpy/sunpy-frontiers-paper}} that accompanies this paper.

\section*{Funding}
\label{sec:funding}

W.T.B., A.Y.S., and S.J.M. are supported by an award from the NASA Research Opportunities in Space and Earth Sciences (ROSES) Open-Source Tools, Frameworks, and Libraries (OSTFL) program.
N.F is supported by NASA under contract NNG09FA40C ({\it IRIS}) and NNG04EA00C (SDO/AIA).
L.A.H is supported by an ESA Research Fellowship.
We acknowledge financial contributions from Google as part of the Google Summer of Code program and from the European Space Agency as part of the Summer of Code in Space program.
We acknowledge financial contributions from NumFOCUS for \enquote{Improving the Usability of \sunpypkg's Data Downloader}.
Additionally, we acknowledge funding from the Solar Physics Division of the American Astronomical Society for \sunpy workshops and tutorials at annual meetings.

\section*{Acknowledgments}
\label{sec:acknowledgments}

We thank everyone who has supported and contributed to the \sunpyproj in any manner.
\sunpypkg makes use of \astropypkg, a community-developed core Python package for Astronomy \citep{astropy_collaboration_astropy_2022}.

\bibliographystyle{Frontiers-Harvard}
\bibliography{main}


\end{document}